\documentclass[12pt,eqsecnum]{revtex4}
\usepackage{amsfonts}
\usepackage{amssymb}
\usepackage{amsmath}
\usepackage{mathptmx}
\usepackage{bm}

\begin{document}
\title{Resonances of Multichannel Systems\footnote{Based on a talk presented
at the UNISA-JINR Symposium ``Models and Methods in Few- and Many-Body Systems''
(6--9 February 2007, Skukuza, Kruger National Park, South Africa)}}
\author{Alexander K. Motovilov}
\address{Bogoliubov Laboratory of Theoretical Physics,
JINR\\ Joliot-Curie 6, 141980 Dubna, Moscow Region, Russia}


\begin{abstract}
\vspace*{0.5truecm}

We describe structure of the $T$-matrices, scattering matrices, and
Green functions on unphysical energy sheets in multichannel
scattering problems with binary channels and in the three-body
problem. Based on the explicit representations obtained for the
values of $T$- and $S$-matrices on the unphysical sheets, we prove
that the resonances belonging to an unphysical sheet are just those
energies where the correspondingly truncated scattering matrix,
taken in the physical sheet, has eigenvalue zero. We show, in
addition, that eigenvectors of the truncated scattering matrix
associated with its zero eigenvalue are formed of the breakup
amplitudes for the respective resonant states.
\end{abstract}

\maketitle

\section{Introduction}

Resonances of multichannel systems play a crucial role in various
problems of nuclear, atomic, and molecular physics. In a wider
sense, resonances represent one of the most interesting and
intriguing phenomena observed in scattering processes, and not only
in quantum physics but also in optics, acoustics, radiophysics,
mechanics of continua etc. Literature on resonances is enormous and
in this short introduction we have a chance to mention only several
key points in the history of the subject and to refer only to a few
key approaches to quantum-mechanical resonances, necessarily leaving
many others a part.

With a resonance of a quantum system one usually associates an
unstable state that only exists during a certain time. The original
idea of interpreting resonances in quantum mechanics as complex
poles of the scattering amplitude (and hence, as those of the
scattering matrix) goes back to G.\,Gamov \cite{GGamow}. For
radially symmetric potentials, the interpretation of two-body
resonances as poles of the analytic continuation of the scattering
matrix has been entirely elaborated in terms of the Jost functions
\cite{Jost}. Beginning with E.\,C.\,Titchmarsh \cite{Titchmarsh} it
was also realized that the $S$-matrix resonances may show up as
poles of the analytically continued Green functions.

Another, somewhat distinct approach to resonances is known as the
complex scaling (or complex rotation) method. The complex scaling
makes it possible to rotate the continuous spectrum of the $N$-body
Hamiltonian in such a way that resonances in certain sectors of the
complex energy plane turn into usual eigenvalues of the scaled
Hamiltonian. In physics literature the origins of such an approach
are traced back at least to C.\,Lovelace \cite{Lovelace1964}. A
rigorous approval of the complex scaling method has been done by
E.\,Balslev and J.\,M.\,Combes \cite{BalslevCombes}. A link between
the $S$-matrix interpretation of resonances and its complex rotation
counterpart was established by G.\,A.\,Hagedorn \cite{Hagedorn} who
has proven that for a reasonable class of quickly decreasing
potentials at least a part of the scaling resonances for an $N$-body
system ($N\leq 4$) turns to be also the scattering matrix
resonances. We remark that the complex scaling seems to be the most
popular approach to practical calculation of resonances,
particularly in atomic and molecular systems (see, e.g., Refs.
\cite{BrandasElander,FedorovGJ-2003,Hu,Korobov,Moiseyev} and
references cited therein).

If support of the interaction is compact, the resonances of a
two-body system can be treated within the approach created by
P.\,Lax and R.\,Phillips \cite{LaxP}. An advantage of the
Lax-Phillips approach is in the opportunity of giving an elegant
operator interpretation of resonances. The two-body
resonances show up as the discrete spectrum of a dissipative
operator which is the generator of the compressed evolution
semigroup. An operator interpretation of resonances in multichannel
systems, based on a $2\times2$ operator matrix representation of a
rather generic Hamiltonian, can be found in \cite{MennMot}.

For more details on the history of the subject and other approaches
to resonances, as well as for the bibliography we refer to books
\cite{AlfaroRegge,Baz,BohmQM,Newton,ReedSimonIII,ReedSimonIV} (it
might also be useful to look through the review parts of papers
\cite{MoellerOrlov} and \cite{MN1997}). Here we only notice that, in
contrast to the ``normal'' bound and scattering states, the resonant
ones still remain a quite mysterious object and many questions
related to resonances are still unanswered. This is partly related
to the fact that, unlike the ``normal'' spectrum, resonances
are not a unitary invariant of a self-adjoint (Hermitian) operator.
Moreover, following to J. S. Howland \cite{Howland1974} and
B.\,Si\-mon~\cite{Simon}, one should conclude that no satisfactory
definition of resonance can rely on a single operator on an abstract
Hilbert space and always an extra structure is necessary. Say, an
unperturbed dynamics (in quantum scattering theory) or geometric
setup (in acoustical or optical problems). Resonances are as
relative as the scattering matrix is itself.

In our approach present in this report we follow the typical setup
where the resonances arising due to an interaction $V$ are
considered relative to the unperturbed dynamics described by the
kinetic energy operator $H_0$. The resolvent $G(z)=(H-z)^{-1}$ of
the total Hamiltonian $H=H_0+V$ is an analytic operator-valued
function of $z\in\mathbb{C}\setminus\sigma(H)$. The spectrum
$\sigma(H)$ of $H$ is a natural boundary for holomorphy domain of
$G(z)$ considered as an operator-valued function.  However the
kernel $G(\cdot,\cdot,z)$ may admit analytic continuation through
the continuous spectrum of $H$. Or the form $\langle
G(z)\varphi,\psi\rangle$ may do this for any $\varphi,\psi$ of a
dense subset of the Hilbert space $\mathfrak{H}$. Or the
``augmented'' resolvent $PG(z)P$ admits such a continuation for $P$
the orthogonal projection onto a subspace of $\mathfrak{H}$. In any
of these cases one deals with the Riemann surface of an analytical
function.

In the simplest example with $H=H_0=-\Delta$, the two-body kinetic
energy operator in coordinate representation, we have
$$
G({\bm{x}},{\bm{x}}',z)=\frac{1}{4\pi}\frac{\mathrm{e}^{\mathrm{i}
z^{1/2}|{\bm{x}}-{\bm{x}}'|}}{|{\bm{x}}-{\bm{x}}'|},
$$
where ${\bm{x}},{\bm{x}}'$ are three-dimensional vectors. Clearly,
$G({\bm{x}},{\bm{x}}',z)$ as a function of the energy $z$ has a
two-sheeted Riemann surface which simply coincides with that of the
function $z^{1/2}$.

In this way one arrives at the concept of the unphysical
energy sheet(s). The copy of the complex energy plane where the
resolvent $G(z)$ is considered initially as an operator-valued
function is called the physical sheet. The remainder of the Riemann
surface is assumed to consist of the unphysical sheets (in general,
an unphysical sheet may only be a small part of the complex plane).

Meanwhile, any analytic function is uniquely defined by its values
given for an infinite set of points belonging to its initial domain
and having at list one limiting point. Usually one knows the
$T$-matrix or Green function on the whole physical sheet which means
that, at least in principle, it should be possible to express their
values on unphysical sheets through the ones on the physical sheet.

In \cite{MN1997,TMF1993} (see also \cite{Mot-ECaYa-2001,MotDSc}) we
have found just such expressions. More precisely, we have derived
explicit representations for the values of the two- and three-body
$G(z)$, $T(z)$, and $S(z)$ on unphysical energy sheets in terms of
these quantities themselves only taken on the physical sheet. The
same has been also done for analogous objects in multichannel
scattering problems with binary channels \cite{TMF1993}. The
representations obtained not only disclose the structure
$T$-matrices, scattering matrices, and Green functions on unphysical
energy sheets but they also show which blocks of the scattering
matrix taken in the physical sheet are ``responsible'' for
resonances on a certain unphysical sheet. This result paves the way
to developing new methods for practical calculation of resonances in
concrete multichannel systems and, in particular, in the three-body
ones (see, e.g. \cite{CPC2000,YaF1997,YaF1999}). As a matter of fact
we reduce all the study of resonances to a work completely on the
physical sheet.

The present report essentially extends the presentation given
recently in \cite{Mot2006}.

\section{Two-body problem}
\label{Sec-2-body}

In general, we assume that the interaction potential $v$ falls off
in coordinate space not slower than exponentially. When studying
resonances of a two--body system with such an interaction one can
employ equally well both coordinate and momentum representations.
However in the three-body case it is much easier for us to work in
the momentum space (for an explanation see \cite{MN1997}, p. 149).
This is one of the reasons why we proceed in the same way in the
two-body case. Thus, for the two-body kinetic energy operator $h_0$
we set $(h_0 f)({\bm{k}})={\bm{k}}^2 f({\bm{k}})$ where
${\bm{k}}\in\mathbb{R}^3$ stands for the reduced relative momentum.
In case of a local potential we have
$v({\bm{k}},{\bm{k}}')=v({\bm{k}}-{\bm{k}}')$ and
$v({\bm{k}})=\overline{v(-{\bm{k}})}$.

Surely, we need to add some requirements on the analyticity of the
potentials in their complex momentum variables as well as on their
fall-off as the real parts of the momenta approach infinity (see
\cite{MN1997} and \cite{TMF1993} for details). For simplicity,
through all this presentation we assume that the potentials involved
are holomorphic functions of the momenta on the corresponding whole
complex spaces (that is, on the whole $\mathbb{C}^3$ in the two-body
case).

The transition operator ($T$-matrix) reads
\begin{equation}
\label{T} t(z)=v-vg(z)v,
\end{equation}
where $g(z)=(h-z)^{-1}$ denotes the resolvent of the perturbed
Hamiltonian $h=h_0+v$. The operator $t$ is the solution of the
Lippmann-Schwinger equation
\begin{equation}
\label{LScht2} t(z)=v-vg_0(z)t(z),
\end{equation}
that is, in terms of the its kernel we have
\begin{equation}
\label{LSh1} t({\bm{k}},{\bm{k}}',z)=v({\bm{k}},{\bm{k}}')-
\int_{\mathbb{R}^3}d{\bm{q}}
\dfrac{v({\bm{k}},{\bm{q}})t({\bm{q}},{\bm{k}}',z)}{{\bm{q}}^2-z}
\end{equation}
taking into account that the free Green function $g_0$ reads
$$
g_0({\bm{k}},{\bm{k}}',z)=
\dfrac{\delta({\bm{k}}-{\bm{k}}')}{({\bm{k}}^2-z)}.
$$

Clearly, all dependence of $t$ on $z$ is determined by the integral
term  on the right-hand side of \eqref{LSh1} that looks like a
particular case of the Cauchy type integral
\begin{equation}
\label{Phi}
\Phi(z)=\int_{\mathbb{R}^N}d{\bm{q}}\dfrac{f({\bm{q}})}{\lambda+
{\bm{q}}^2-z}
\end{equation}
for $N=3$. Cauchy type integrals of the same form but for various
$N$ we will also have below when considering a multichannel problem
with binary channels in Sec. \ref{Sec-mult} and the three-body
problem in Sec. \ref{Sec-3-body}.

Let $\mathfrak{R}_\lambda$, $\lambda\in\mathbb{C}$, be the Riemann
surface of the function
\begin{equation}
\label{zeta} \zeta(z)=\left\{\begin{array}{ll}
(z-\lambda)^{1/2} & \text{if $N$ is odd}, \\
\log(z-\lambda) & \text{if $N$ is even}.
\end{array}
\right.
\end{equation}
If $N$ is odd, $\mathfrak{R}_\lambda$ is formed of two sheets of the
complex plane. One of them, where $(z-\lambda)^{1/2}$ coincides with
the arithmetic square root $\sqrt{z-\lambda}$, we denote by $\Pi_0$.
The other one, where $(z-\lambda)^{1/2}=-\sqrt{z-\lambda}$, is
denoted by $\Pi_1$.

If $N$ is even, the number of sheets of $\mathfrak{R}_\lambda$ is
infinite. In this case as the index $\ell$ of a sheet $\Pi_\ell$ we
take the branch number of the function $\log(z-\lambda)$ picked up
from the representation
$\log(z-\lambda)=\log|z-\lambda|+\mathrm{i}\,2\pi\ell
+\mathrm{i}\phi$ with $\phi\in[0,2\pi)$.

Usually the point $\lambda$ is called the branching point of the
Riemann surface $\mathfrak{R}_\lambda$.

The following statement can be easily proven by applying the residue
theorem (if necessary, consult \cite{TMF1993} for a proof).
\smallskip

\noindent{\bf Lemma 1}. {\it For a holomorphic $f({\bm{q}})$,
${\bm{q}}\in\mathbb{C}^N$, the function $\Phi(z)$ given by
\eqref{Phi} is holomorphic on $\mathbb{C}\setminus[\lambda,
+\infty)$ and admits the analytic continuation onto the unphysical
sheets $\Pi_\ell$ of the Riemann surface $\mathfrak{R}_\lambda$ as
follows
\begin{equation}
\label{Phi1} \Phi(z|_{\Pi_{\ell}})=\Phi(z)-\ell\,\pi\mathrm{i}
(\sqrt{z-\lambda}\,\,)^{N-2} \int_{S^{N-1}}d\widehat{{\bm{q}}}\,
f(\sqrt{z-\lambda}{\widehat{\bm{q}}}),
\end{equation}
where $S^{N-1}$ denotes the unit sphere in $\mathbb{R}^N$ centered
at the origin. }

Notice that in \eqref{Phi1} and further on the writing
$z\bigl|_{\Pi_\ell}$ means that position of $z$ is taken on the
unphysical sheet $\Pi_\ell$. If the reference to $\Pi_\ell$ is not
present and we write simply $z$ than one understands that we deal
with exactly the same energy point but lying on (dropped onto) the
physical sheet $\Pi_0$.

Now return to the two-body problem and set
$$
\bigl(g_0(z)f_1,f_2\bigr)\equiv \int_{\mathbb{R}^3}d{\bm{q}}\,
\frac{f_1({\bm{q}})f_2({\bm{q}})}{{\bm{q}}^2-z},
$$
where $f_1$ and $f_2$ are holomor\-phic. Then by Lem\-ma~1
$$
\bigl(g_0(z|_{\Pi_1})f_1,f_2\bigr)=\bigl(g_0(z|_{\Pi_0})f_1,f_2\bigr)-
\pi\mathrm{i}\sqrt{z}\int_{S^{2}}d\widehat{{\bm{q}}}\,
f_1(\sqrt{z}{\widehat{\bm{q}}})f_2(\sqrt{z}{\widehat{\bm{q}}}),
$$
which means that the continuation of the free Green function
$g_0(z)$ onto the unphysical sheet $\Pi_1$ can be written in short
form as
\begin{equation}
\label{g0c} g_0(z|_{\Pi_1})=g_0(z)+\mathrm{a}_0(z)j^\dagger(z)j(z),
\end{equation}
where $\mathrm{a}_0(z)=-\pi\mathrm{i}\sqrt{z}$ and $j(z)$ is the
operator forcing a (holomorphic) function $f$ to set onto the energy
shell, i.e.
$\bigl(j(z)f\bigr)({{\widehat{\bm{k}}}})
=f(\sqrt{z}{\widehat{\bm{k}}})$.

Taking into account \eqref{g0c}, on the unphysical sheet $\Pi_1$ the
Lippmann-Schwinger equation \eqref{LScht2} turns into
$$
t'=v-v(g_0+\mathrm{a}_0{j^\dagger}j) t', \qquad t'=t|_{\Pi_1}.
$$
Hence $ (I+vg_0)t'=v-\mathrm{a}_0{j^\dagger}j\, t'.$ Invert $I+vg_0$
by using the fact that $t(z)=v-vg_0t$ and, hence,
$(I+vg_0)^{-1}v=t$:
\begin{equation}
\label{t1} t'=t-\mathrm{a}_0 t{j^\dagger}j t'.
\end{equation}
Apply $j(z)$ to both sides of \eqref{t1} and obtain $j t'=j
t-\mathrm{a}_0\, j t{j^\dagger}\,j t'$,  which means
\begin{equation}
\label{t11} (\widehat{I}+\mathrm{a}_0\, j t{j^\dagger})\,j t'=j t,
\end{equation}
where $\widehat{I}$ stands for the identity operator in $L_2(S^2)$.
Then observe that $\widehat{I}+\mathrm{a}_0\, j t{j^\dagger}$ is
nothing but the two-body scattering matrix $s(z)$ since the kernel
of the latter for $z\in\Pi_0$ is known to read
$$
s({\widehat{\bm{k}}},{\widehat{\bm{k}}}',z)=
\delta({\widehat{\bm{k}}},{\widehat{\bm{k}}}')-\pi\mathrm{i}\sqrt{z}
\,\,t(\sqrt{z}{\widehat{\bm{k}}},\sqrt{z}{\widehat{\bm{k}}}',z).
$$
Hence
\begin{equation}
\label{jtpst} j t'=[s(z)]^{-1}j t.
\end{equation}
Now go back to \eqref{t1} and by using \eqref{jtpst} get
$t'=t-\mathrm{a}_0 \,\,t{j^\dagger} [s(z)]^{-1} j t$, that is,
\begin{equation}
\label{t11p}  t(z|_{\Pi_1})=t(z)-\mathrm{a}_0(z)
\,\,t(z){j^\dagger}(z)[s(z)]^{-1}\,j(z)t(z).
\end{equation}
All entries on the right-hand side of \eqref{t11p} are on the
physical sheet. This is just the representation for the two-body
$T$-matrix on the unphysical sheet we looked for.

{}From \eqref{t11p} one immediately derives representations for the
continued resolvent,%
\begin{equation}
\label{gr}
 g(z|_{\Pi_1})=g+\mathrm{a}_0\,(I-gv){j^\dagger}\,[s(z)]^{-1}j(I-vg),
\end{equation}
and continued scattering matrix,
\begin{equation}
\label{sr} s(z|_{\Pi_1})=\mathcal{E}\,[s(z)]^{-1}\,\mathcal{E},
\end{equation}
where $\mathcal{E}$ is the inversion,
$(\mathcal{E}f)({\widehat{\bm{k}}})=f(-{\widehat{\bm{k}}})$. Hence,
the resonances are nothing but zeros of the scattering matrix $s(z)$
in the physical sheet. That is, the energy $z$ on the unphysical
sheet $\Pi_1$ is a resonance if and only if there is a non-zero
vector $\mathcal{A}$ of $L_2(S^2)$ such that
\begin{equation}
s(z)\mathcal{A}=0 \label{s2bodyA}
\end{equation}
for the same $z$ on the physical sheet.

We remark that this fact is rather well known for the partial-wave
Schr\"odinger equations in case of centrally-symmetric potentials.
For this case the statement that the resonances correspond to zeros
of the partial-wave scattering matrix $s_{l}$ on the physical sheet
of the com\-p\-lex energy plane follows from its representation (see,
e.g., \cite{AlfaroRegge})
$$
s_{l}(p)=(-1)^{l}\dfrac{f_l(p)}{f_l(-p)}
$$
in terms of the Jost function $f_l(p)$ where $l$ stands for the
angular momentum and $p$ for the (scalar) complex momentum. This
property of $s_{l}(p)$ was explicitly noticed in the
review article \cite[p.\,1357]{MoellerOrlov}. Generalizations of the
statement to the case of multichannel problems with binary channels
and to the three-body problem have been given in \cite{TMF1993} and
\cite{MN1997}, respectively. We will discuss them below in Sec.
\ref{Sec-mult} and \ref{Sec-3-body}.

The eigenfunction $\mathcal{A}$ in \eqref{s2bodyA} represents
the breakup amplitude of an unstable state associated with the
resonance $z$.  This means that in coordinate space the
corresponding ``Ga\-mov vector'', i.e. the resonance solution to the
Schr\"odinger equation, has the following asymptotics
\begin{align}
\label{3b-ras}
\psi_{\mathrm{res}}({\bm{x}})\mathop{\sim}\limits_{{\bm{x}}\to\infty}&
\mathcal{A}(-\widehat{{\bm{x}}})
\dfrac{\exp\bigl(\mathrm{i} z^{1/2}\bigl|_{\Pi_1}
|{\bm{x}}|\bigr)}{|{\bm{x}}|}\\
\nonumber
&\qquad\qquad=\mathcal{A}(-\widehat{{\bm{x}}})
\dfrac{\mathrm{e}^{-\mathrm{i}\sqrt{z}|{\bm{x}}|}}{|{\bm{x}}|},
\qquad \widehat{{\bm{x}}}={{\bm{x}}}/{|{\bm{x}}|}.
\end{align}
This claim is a particular case of the statement of Lemma 2 below.

It should be stressed that the asymptotics \eqref{3b-ras} contains
no term with the incoming spherical wave
$$
\dfrac{\exp\bigl(-\mathrm{i}
z^{1/2}\bigl|_{\Pi_1}|{\bm{x}}|\bigr)}{|{\bm{x}}|}.
$$

We conclude the section with a remark that in \cite{Orlov1984} (see
also \cite[Section 2]{OrlovTur} and
\cite[Sec\-ti\-on~3]{MoellerOrlov}) Yu.\,V. Orlov was very close to
obtaining a representation that would be a version of the
representation \eqref{t11p} for partial-wave two-body $T$-matrices
in the case of centrally-symmetric potentials. As a matter of fact,
only the last step has not been done in \cite{Orlov1984,OrlovTur},
the one analogous to the transition from equation \eqref{t1} to
equation \eqref{t11p} by using relation~\eqref{jtpst}.

\section{Multichannel problem with binary channels}
\label{Sec-mult}

From now on assume that $h$ is an $m\times m$ matrix Schr\"odinger
operator of the form
\begin{equation}
\label{h-mult}
h=\left(\begin{array}{cccc}
\lambda_1+h_0^{(1)}+v_{11} & {v}_{12} & \ldots & {v}_{1m} \\
{v}_{21} & \lambda_2+h_0^{(2)}+{v}_{22} & \,\,\ldots\,\, & {v}_{2m}\\
\ldots & \ldots &  \ldots & \ldots \\
{v}_{m1} & {v}_{m2}&   \ldots &  \lambda_m+h_0^{(m)}+{v}_{mm}
\end{array}\right),
\end{equation}
written in the momentum representation. Thus, we assume that
$$
(h^{(\alpha )}_0 f_\alpha )({\bm{k}}_\alpha )= {\bm{k}}_\alpha ^2
f_\alpha ({\bm{k}}_\alpha ), \quad {\bm{k}}_\alpha \in
{\mathbb{R}}^{n_\alpha }, \quad f_\alpha \in L_2
({\mathbb{R}}^{n_\alpha }), \quad \alpha=1,2,\ldots,m.
$$
We restrict ourselves to the case where the channel dimensions
$n_\alpha$ satisfy inequalities $n_\alpha\geq 3$,
$\alpha=1,2,\ldots,m$, and $1\leq m<\infty$. For simplicity we
assume that the potential/coupling terms
${v}_{\alpha\beta}({\bm{k}}_\alpha,{\bm{k}}'_\beta)$ are holomorphic
functions of their variables
${\bm{k}}_\alpha\in{\mathbb{C}}^{n_\alpha}$ and
${\bm{k}}'_\beta\in{\mathbb{C}}^{n_\beta}$, sufficiently rapidly
decreasing as $\mathop{\mathrm{Re}} {\bm{k}}_\alpha\to\infty$ or
$\mathop{\mathrm{Re}} {\bm{k}}'_\beta\to\infty$ (see
\cite{TMF1993}). The thresholds
$\lambda_1,\lambda_2,\ldots,\lambda_m\in\mathbb{R}$ are supposed to
be distinct and arranged in ascending order:
$\lambda_1<\lambda_2<\ldots<\lambda_m$.

We also introduce the notations
$$
h_0=\left( \begin{array}{cccc}
\lambda_1+h_0^{(1)} & 0 & \ldots & 0 \\
0 & \lambda_2+h_0^{(2)} & \ldots & 0 \\
\ldots & \ldots & \ldots & \ldots \\
0 & 0 & \ldots & \lambda_m+h_0^{(m)}
\end{array}\right )
\quad\text{and}\quad {v}=\left( \begin{array}{cccc}
{v}_{11} & {v}_{12} &\ldots & {v}_{1m}\\
{v}_{21} & {v}_{22} &\ldots & {v}_{2m}\\
{\ldots} &\ldots & \ldots & \ldots \\
{v}_{m1} & {v}_{m2} &\ldots & {v}_{mm} \end{array}\right )
$$
for the unperturbed Hamiltonian and the total interaction,
respectively. By $g_0(z)$ and $g(z)$ we denote the corresponding
resolvents,
$$
g_0(z)=(h_0-z)^{-1}\quad\text{and}\quad g(z)=(h-z)^{-1}.
$$

Like in the one-channel (i.e. two-body) case of
Sec.~\ref{Sec-2-body}, we again begin with the study of the
\textit{T}-matrix
$$
t(z)={v}-{v}g(z){v}
$$
that is the solution to the Lippman-Schwinger equation
\begin{equation}
    t(z)={v}-{v}g_0(z)t(z).
\label{LSchMult}
\end{equation}
Kernels $t_{\alpha\beta}({\bm{k}}_\alpha,{\bm{k}}'_\beta,z)$ of the
block entries $t_{\alpha\beta}(z)$  of the operator matrix $t(z)$
solve the equation system
\begin{equation}
\label{Pap1-3}
t_{\alpha\beta}({\bm{k}},{\bm{k}}',z)=
{v}_{\alpha\beta}({\bm{k}},{\bm{k}}')-
\sum\limits_{\gamma=1}^{m}\int_{\mathbb{R}_{n_\gamma}}
d{\bm{q}}\dfrac{{v}_{\alpha\gamma}({\bm{k}},{\bm{q}})
t_{\gamma\beta}({\bm{q}},{\bm{k}}',z)}
{\lambda_\gamma+{\bm{q}}^2-z}.
\end{equation}
Similarly to the two-body $T$-matrix in equation \eqref{LSh1}, all
dependence of the kernels $t_{\alpha\beta}({\bm{k}},{\bm{k}}',z)$ on
$z$ is determined by the integrals on the right-hand-side of
(\ref{Pap1-3}), which are the Cauchy type integrals  just of the
form \eqref{Phi}.

In contrast to the two-body case, for $m\geq 2$ we arrive at a
multi-sheeted Riemann surface with number of sheets greater than two.
The reason is simple: every threshold $\lambda_\alpha$,
$\alpha=1,2,\ldots,m$, turns into a branching point. If all the
channel dimensions $n_\alpha$ are odd, the number of sheets should
be equal to $2^m$, that is, in addition to the physical sheet the
Riemann surface will contain $2^{m}-1$ unphysical ones. If at least
one of $n_\alpha$'s is even, we will have a logarithmic branching
point and the number of unphysical sheets will be necessarily
infinite. In fact, this Riemann surface simply coincides with the
Riemann surface $\mathfrak{R}$ of the vector-valued function
$$
\boldsymbol{\zeta}(z)=\bigl(\zeta_1(z),\zeta_2(z),
\ldots,\zeta_m(z)\bigr),
$$
where (cf. formula \eqref{zeta})
\begin{align*}
\zeta_\alpha(z)&=\left\{\begin{array}{ll}
(z-\lambda_\alpha)^{1/2} & \quad \text{if $n_\alpha$ is odd},\\
\log(z-\lambda_\alpha) & \quad \text{if $n_\alpha$ is even},
\end{array}
\right.\\
&\qquad \alpha=1,2,\ldots,m.
\end{align*}
To enumerate the sheets of $\mathfrak{R}$ it is natural to use a
multi-index
$$
\ell=(\ell_1,\ell_2,\ldots,\ell_m),
$$
where each $\ell_\alpha$ coincides with the branch number for the
corresponding function $\zeta_\alpha$, $\alpha=1,2,...,m$. In
particular, if $n_\alpha$ is odd then $\ell_\alpha$ may get only two
values: either 0 or 1. For even $n_\alpha$ the value of
$\ell_\alpha$ is allowed to be any integer. The sheets of
$\mathfrak{R}$ are denoted by $\Pi_\ell$. The physical sheet
corresponds to the case where all components of $\ell$ are equal to
zero and thus it is denoted simply by $\Pi_0$.

Each sheet $\Pi_\ell$ is a copy of the complex plane ${\mathbb{C}}'$
cut along the ray $[\lambda_1,+\infty)$. The sheets are pasted to
each other in a suitable way along edges of the cut segments between
neighboring points in the set of the thresholds $\lambda_\alpha$,
$\alpha =1,2,\ldots,m$. In particular, if coming from the sheet
$\Pi_{(\ell_1,\ell_2,\ldots,\ell_m)}$ the energy $z$ crosses the
interval $(\lambda_\alpha,\lambda_{\alpha+1})$,
$\alpha=1,2,\ldots,m$, $\lambda_{m+1}\equiv+\infty$, in the upward
direction (i.e. passes from the region $\mathop{\mathrm{Im}} z<0$ to
the region $\mathop{\mathrm{Im}} z>0$), then it arrives at the sheet
$\Pi_{(\ell'_1,\ell'_2,\ldots,\ell'_\alpha,
\ell_{\alpha+1},\ldots,\ell_m)}$ with all indices beginning from
$\ell_{\alpha+1}$ remaining the same while the first $\alpha$
indices $\ell_j$, $1\leq j\leq\alpha$, change by unity. If $n_j$ is
odd then $\ell'_j=1$ for $\ell_j=0$ and $\ell'_j=0$ for $\ell_j=1$;
if $n_j$ is even then $\ell'_j=\ell_j+1$. In the case where the
energy $z$ passes the same interval
$(\lambda_\alpha,\lambda_{\alpha+1})$ downward, it arrives at the
sheet $\Pi_{(\ell'_1,\ell'_2,\ldots,\ell'_\alpha,
\ell_{\alpha+1},\ldots,\ell_m)}$ where for odd $n_j$ the indices
$\ell'_j$ are the same as in the previous case and for even $n_j$
they change according to the rule $\ell'_j=\ell_j-1$. The indices
$\ell_j$ with numbers $j\geq\alpha+1$ remain unchanged.

Under the assumption that the kernels
$t_{\alpha\beta}(\sqrt{z-\lambda_\alpha}
{\widehat{\bm{k}}},{\bm{k}}',z)$ admit the analytic continuation in
$z$ through the cuts (the existence of such a continuation may be
rigorously approved, see \cite{TMF1993}) one can perform analytic
continuation of the Lippman-Schwinger equation \eqref{Pap1-3} from
the physical sheet $\Pi_0$ onto any unphysical sheet $\Pi_\ell$ of
the surface $\mathfrak{R}$. Of course, the trajectory along which we
pull $z$ should avoid the branching points $\lambda_\alpha$.
Applying after each crossing the interval $(\lambda_1,+\infty)$ the
corresponding variant of formula \eqref{Phi1} we arrive at
the following result
\begin{align}
\label{tPil}
t_{\alpha\beta}({\bm{k}},{\bm{k}}',z\bigl|_{\Pi_\ell})=&
{v}_{\alpha\beta}({\bm{k}},{\bm{k}}')-
\sum\limits_{\gamma=1}^{m}\int_{\mathbb{R}_{n_\gamma}}
d{\bm{q}}\dfrac{{v}_{\alpha\gamma}({\bm{k}},{\bm{q}})
t_{\gamma\beta}({\bm{q}},{\bm{k}}',z\bigl|_{\Pi_\ell})}
{\lambda_\gamma+{\bm{q}}^2-z}\\
\nonumber &-\sum\limits_{\gamma=1}^{m}\,\,\ell_\gamma\, A_\beta(z)
\int_{S^{n_\gamma-1}}d\widehat{{\bm{q}}}\,\,\,{v}_{\alpha\gamma}
({\bm{k}}_\alpha,\sqrt{z-\lambda_\gamma}\widehat{{\bm{q}}})
\, t_{\gamma\beta}(\sqrt{z-\lambda_\gamma}
\widehat{{\bm{q}}},{\bm{k}}'_\beta,z\bigl|_{\Pi_\ell})\,,
\end{align}
where
\begin{equation}
\label{Pap1-Af}
A_\gamma(z)=-\pi\mathrm{i}(\sqrt{z-\lambda_\gamma})^{n_\gamma-2}
\end{equation}
Notice that the second integral term on the right-hand side of
\eqref{tPil} includes the half-on-shell values
$t_{\gamma\beta}(\sqrt{z-\lambda_\gamma}
\widehat{{\bm{q}}},{\bm{k}}'_\beta,z\bigl|_{\Pi_\ell})$ of the
$T$-matrix kernels
$t_{\gamma\beta}({\bm{q}},{\bm{k}}'_\beta,z\bigl|_{\Pi_\ell})$ taken
on the unphysical sheet $\Pi_\ell$. Thus, like in the two-body case
of Sec. \ref{Sec-2-body}, it is convenient to introduce operators
$j_\gamma(z)$ forcing a holomorphic function $f({\bm{q}})$,
${\bm{q}}\in\mathbb{C}^n$, to set onto the corresponding energy
shell, i.e.
$$
\bigl(j_\gamma(z)f\bigr)(\widehat{{\bm{q}}})=
f(\sqrt{z-\lambda_\gamma}\widehat{{\bm{q}}}), \quad
\gamma=1,2,\ldots,m.
$$
From these operators we construct a block diagonal matrix
$$
J(z)=\left( \begin{array}{cccc}
j_{1}(z) & 0 &\ldots & 0\\
0 & j_{2}(z) &\ldots & 0\\
{\ldots} &\ldots & \ldots & \ldots \\
0 & 0 &\ldots & j_{m}(z) \end{array}\right ).
$$
Using this notation one easily rewrites equation \eqref{tPil} in the
matrix form
\begin{equation}
\label{tPilM} t(z\bigl|_{\Pi_\ell})=v
-vg_0(z)t(z\bigl|_{\Pi_\ell})-vJ^\dagger(z)LA(z)J(z)t(z\bigl|_{\Pi_\ell}),
\end{equation}
where $L$ and $A(z)$ are diagonal $m\times m$ matrices with scalar
entries,
\begin{equation}
\label{LAs} L=\left( \begin{array}{cccc}
\ell_{1} & 0 &\ldots & 0\\
0 & \ell_{2} &\ldots & 0\\
{\ldots} &\ldots & \ldots & \ldots \\
0 & 0 &\ldots & \ell_{m} \end{array}\right )\quad\text{and}\quad
A(z)=\left( \begin{array}{cccc}
A_{1}(z) & 0 &\ldots & 0\\
0 & A_{2}(z) &\ldots & 0\\
{\ldots} &\ldots & \ldots & \ldots \\
0 & 0 &\ldots & A_{m}(z) \end{array}\right ),
\end{equation}
and $J^\dagger(z)$ is the ``transpose'' of $J(z)$ which means that
the product $t(z)J^\dagger(z)$ has half-on-shell kernels of the form
$v_{\alpha\beta}({\bm{k}},\sqrt{z-\lambda_\beta}\widehat{{\bm{k}}}')$.

When rearranging \eqref{tPilM} we first transfer the term
$vg_0(z)t(z\bigl|_{\Pi_\ell})$ to the left-hand side of
\eqref{tPilM} and obtain
\begin{equation}
\label{Ivg} \bigl(I+vg_0(z)\bigr)t(z\bigl|_{\Pi_\ell})=
v-vJ^\dagger(z)LA(z)J(z)t(z\bigl|_{\Pi_\ell}).
\end{equation}
Our next step is to invert the operator $\bigl(I+vg_0(z)\bigr)$ (of
course, this is only possible for $z$ not belonging to the discrete
spectrum of $h$). Here, we keep in mind that the energy $z$ in this
operator is from the physical sheet where the Lippmann-Schwinger
equation \eqref{LSchMult} holds and thus
$\bigl(I+vg_0(z)\bigr)^{-1}v=t(z)$. Using this inversion formula we
then derive from \eqref{Ivg} that
\begin{equation}
\label{Ivg1} t(z\bigl|_{\Pi_\ell})=
t(z)-t(z)J^\dagger(z)LA(z)J(z)t(z\bigl|_{\Pi_\ell}).
\end{equation}
At this point it is convenient to introduce another diagonal scalar
$m\times m$ matrix
\begin{equation}
\label{tL} \widetilde{L}=\left( \begin{array}{cccc}
\widetilde{\ell}_{1} & 0 &\ldots & 0\\
0 & \widetilde{\ell}_{2} &\ldots & 0\\
{\ldots} &\ldots & \ldots & \ldots \\
0 & 0 &\ldots & \widetilde{\ell}_{m} \end{array}\right )
\end{equation}
whose diagonal entries are
$$
\widetilde{\ell}_\alpha=\left\{\begin{array}{cl}
0 & \quad\text{if $\ell_\alpha=0$}, \\
\mathop{\rm Sign}(\ell_\alpha)=\dfrac{\ell_\alpha}{|\ell_\alpha|}&
\quad\text{if $\ell_\alpha\neq0$.}
\end{array}
\right.
$$
Clearly, the matrices $L$, $\widetilde{L}$, and $A(z)$ commute.
Moreover,
$L\widetilde{L}=L.$
Using these facts one rewrites \eqref{Ivg1} in a slightly different
form
\begin{equation}
\label{Ivg2} t(z\bigl|_{\Pi_\ell})=
t(z)-t(z)J^\dagger(z)LA(z)\widetilde{L}J(z)t(z\bigl|_{\Pi_\ell}).
\end{equation}
which means that the value $t(z\bigl|_{\Pi_\ell})$ of the $T$-matrix
$t$ at a point $z$ on the unphysical sheet $\Pi_\ell$ is expressed
through the value of $t$ itself taken at the same point $z$ on the
physical sheet as well as through the half-on-shell value
$J(z)t(z\bigl|_{\Pi_\ell})$ taken still for $z\bigl|_{\Pi_\ell}$ and, in
addition, multiplied by $\widetilde{L}$ from the left. Applying the
product $\widetilde{L}J(z)$ to both side of \eqref{Ivg2} we arrive
at a closed equation  for $\widetilde{L}J(z)t(z\bigl|_{\Pi_\ell})$:
\begin{equation}
\label{Igv3}
\bigl[\widehat{I}+\widetilde{L}J(z)t(z)J^\dagger(z)LA(z)\bigr]
\widetilde{L}J(z)t(z\bigl|_{\Pi_\ell})=\widetilde{L}J(z)t(z),
\end{equation}
where $\widehat{I}$ denotes the identity operator in the sum Hilbert
space
\begin{equation}
\label{fG} \mathfrak{G}=L_2(S^{n_1-1})\oplus
L_2(S^{n_2-1})\oplus\ldots\oplus L_2(S^{n_m-1}).
\end{equation}
Therefore, at any point $z$ in the physical sheet where the operator
\begin{equation}
\label{sell}
s_\ell(z)=\widehat{I}+\widetilde{L}J(z)t(z)J^\dagger(z)LA(z)
\end{equation}
is invertible, we will have
\begin{equation}
\label{Igv4}
\widetilde{L}J(z)t(z\bigl|_{\Pi_\ell})=
[s_\ell(z)]^{-1}\widetilde{L}J(z)t(z).
\end{equation}
Notice that $s_\ell(z)$ commutes with $\widetilde{L}$, i.e.
$$
\widetilde{L}s_\ell(z)=s_\ell(z)\widetilde{L},
$$
and hence
\begin{equation}
\label{Igv5} LA(z)s_\ell(z)^{-1}\widetilde{L}=LA(z)s_\ell(z)^{-1}.
\end{equation}
Taking into account equalities \eqref{Igv4} and \eqref{Igv5} we
obtain from \eqref{Igv3} the following result:
\begin{align}
\label{Pap1-13}
t\bigl(z\bigl|_{\Pi_\ell}\bigr)&
=t(z)-t(z)J^\dagger(z)LA(z)[{s_\ell}(z)]^{-1}
\widetilde{L} J(z)t(z)\\
\label{Pap1-13p} &=t(z)-t(z)J^\dagger(z)L
A(z)[{s_\ell}(z)]^{-1}(z)J(z)t(z).
\end{align}
These are just the representations for
$t\bigl(z\bigl|_{\Pi_\ell}\bigr)$ we look for: in \eqref{Pap1-13}
and \eqref{Pap1-13p} values of the multichannel $T$-matrix on an
arbitrarily chosen unphysical energy sheet $\Pi_\ell$ are explicitly
written in terms of the entries whose values are taken from the
physical sheet. Formulas \eqref{Pap1-13} and \eqref{Pap1-13p} are
just the ones that represent a generalization of the two-body
representation \eqref{t11p} to the case of multichannel
Schr\"odinger operators with binary channels. A slightly different
version of the representations \eqref{Pap1-13} and  \eqref{Pap1-13p}
was first published in \cite{TMF1993}.

The operator matrix $s_\ell(z)$ given by \eqref{sell} is closely
related to the total scattering matrix for the problem which reads
\begin{equation}
s(z)=\widehat{I}+J(z)t(z)J^\dagger(z)A(z),
\end{equation}
Of course, the total scattering matrix contains neither entry $L$
nor entry $\widetilde{L}$. For the matrix $s_\ell(z)$ these entries
play an important role. Depending on the unphysical sheet $\Pi_\ell$
under consideration, certain rows and columns of the difference
matrix $\bigl(s_\ell(z)-\widehat{I}\,\,\bigr)=
\widetilde{L}J(z)t(z)J^\dagger(z)L$ completely consist of zero
entries. Nullification takes place for those rows and columns of the
difference matrix
$\bigl(s(z)-\widehat{I}\,\,\bigr)=J(z)t(z)J^\dagger(z)$ whose
numbers $\alpha$ are such that the corresponding indices
$\ell_\alpha$ equal zero.  This is a reason why we call $s_\ell(z)$
the truncated scattering matrix associated with the unphysical sheet
$\Pi_\ell$.

Notice that if instead of \eqref{LSchMult} we start with the
transposed Lippmann-Schwinger equation
$$
    t(z)={v}-t(z)g_0(z){v}\,,
$$
then in the same way we obtain for $t\bigl(z\bigl|_{\Pi_\ell}\bigr)$
another representation that can be considered as a transposed
version of the representation (\ref{Pap1-13}):
\begin{align}
\label{Pap1-16}
t\bigl(z\bigl|_{\Pi_\ell}\bigr)&=t(z)-t(z)J^\dagger(z)
\widetilde{L}[s_\ell^\dagger(z)]^{-1}A(z)LJ(z)t(z)\\
\label{Pap1-16p}
&=t(z)-t(z)J^\dagger(z)[s_\ell^\dagger(z)]^{-1}A(z)LJ(z)t(z),
\end{align}
where
\begin{align*}
s_\ell^\dagger(z)=\widehat{I}+
LA(z)J(z)t(z)J^\dagger(z)\widetilde{L}\,.
\end{align*}
The operator $s^\dagger_\ell(z)$ represents the result of truncation
of the transposed \textit{S}-matrix
$$
s^\dagger(z)=\widehat{I}+A(z)J(z)t(z)J^\dagger(z).
$$
From the uniqueness of the analytic continuation by \eqref{Pap1-13}
and \eqref{Pap1-16} it immediately follows that
\begin{align*}
t(z)J^\dagger(z)LA(z)s_\ell(z)^{-1}J(z)t(z)&=
t(z)J^\dagger(z)[s_\ell^\dagger(z)]^{-1}A(z)LJ(z)t(z).
\end{align*}

To describe structure of the scattering matrices $s(z)$ or
$s^\dagger(z)$ analytically continued to an unphysical sheet
$\Pi_\ell$ we need some more notations. First, introduce a block
diagonal operator matrix $\mathcal{E}({\ell})$ of the form
$\mathcal{E}=\mathop{\rm diag}\big({\cal E}_1,{\cal
E}_2,\ldots,{\cal E}_m\big)$ where ${\cal E}_\alpha$ is the identity
operator on $L_2(S^{n_\alpha-1})$ if $\ell_\alpha$ is even and
${\cal E}_\alpha$ is the inversion, $({\cal E}_\alpha
f)(\widehat{{\bm{k}}})=f(-\widehat{{\bm{k}}})$, if $\ell_\alpha$ is
odd. Second, let ${\mathrm{e}}(\ell)$ be a scalar diagonal matrix,
${\mathrm{e}}=\mathop{\rm
diag}\big({\mathrm{e}}_1,{\mathrm{e}}_2,\ldots,{\mathrm{e}}_m\big)$,
with the main diagonal entries ${\mathrm{e}}_\alpha$ defined by
\begin{equation*}
{\mathrm{e}}_\alpha=\left\{
\begin{array}{ll}
+1 & \text{for any $\ell_\alpha=0,\pm1,\pm2,\dots$\,
       if \, } n_\alpha \text{\, is even\,},\\
+1 & \text{if \,} n_\alpha \text{\, is odd and \,} \ell_\alpha=0,\\
-1 & \text{if \,} n_\alpha \text{\, is odd and \,} \ell_\alpha=1.
\end{array}
\right.
\end{equation*}
That is, ${\mathrm{e}}_\alpha$ only depend on the corresponding
$n_\alpha$ and $\ell_\alpha$. It is obvious that if a
matrix-valued function $A(z)$ is defined on the physical sheet of
the Riemann surface $\Re$ by formulas \eqref{Pap1-Af} and
\eqref{LAs}, then after the analytic continuation to the sheet
$\Pi_\ell$ it acquires the form
\begin{align}
\label{Pap1-AzPl} A(z)\big|_{\Pi_\ell}=A(z){\mathrm{e}}(\ell).
\end{align}

Now we are ready to present our main result concerning the
$S$-matrices. We claim that after continuation to the sheet $\Pi_\ell$
their values are expressed by the formulas
\begin{align}
\label{Pap1-18} s\bigl(z\big|_{\Pi_\ell}\bigr)&=\mathcal{E}
\left[\widehat{I}+{{^{\mbox{\scriptsize$\ulcorner$}}}\!\!
t^{\mbox{\scriptsize$\!\urcorner$}}} A\mathrm{e} -
{{^{\mbox{\scriptsize$\ulcorner$}}}\!\!
t^{\mbox{\scriptsize$\!\urcorner$}}} LA
s_\ell^{-1}{{^{\mbox{\scriptsize$\ulcorner$}}}\!\!
t^{\mbox{\scriptsize$\!\urcorner$}}}
A{\mathrm{e}}\right]\mathcal{E}\,,\\
\label{Pap1-19}
s^\dagger\bigl(z\bigl|_{\Pi_\ell}\bigr)&=\mathcal{E}
\left[\widehat{I}+ \mathrm{e}
A{{^{\mbox{\scriptsize$\ulcorner$}}}\!\!
t^{\mbox{\scriptsize$\!\urcorner$}}} - \mathrm{e}
A{{^{\mbox{\scriptsize$\ulcorner$}}}\!\!
t^{\mbox{\scriptsize$\!\urcorner$}}}[s^{\dagger}_\ell]^{-1}
AL\,{{^{\mbox{\scriptsize$\ulcorner$}}}\!\!
t^{\mbox{\scriptsize$\!\urcorner$}}} \right]\mathcal{E}\,,
\end{align}
where we use another shorthand notation
\begin{equation*}
{{^{\mbox{\scriptsize$\ulcorner$}}}\!\!t^{\mbox{\scriptsize$\!
\urcorner$}}}(z)=J(z)t(z){J^\dagger}(z).
\end{equation*}
The argument $z$ of the operator-valued functions $s_\ell(z)$,
$s^\dagger_\ell(z)$, $J(z)$, $J^\dagger(z)$, and $A(z)$ on the
right-hand sides of \eqref{Pap1-18} è \eqref{Pap1-19} is a point on
the physical sheet $\Pi_0$ having just the same position on the
complex plane as the point $z\big|_{\Pi_\ell}$ on the sheet
$\Pi_\ell$ on the left-hand sides of \eqref{Pap1-18} and
\eqref{Pap1-19}, respectively.

At last, we present the representation for the continued resolvent
on the sheet $\Pi_\ell$:
\begin{align}
\label{Pap1-rPil} g\bigl(z\big|_{\Pi_\ell}\bigr)=&g+\big(I-gv
\big)J^\dagger
AL s_\ell^{-1}J\big(I-v g\big),\\
\label{Pap1-rPilt} =&g+\big(I-gv \big)J^\dagger
[s^\dagger_\ell]^{-1}AL J\big(I-v g\big).
\end{align}

In this report we skip derivation of the representations
\eqref{Pap1-18}--\eqref{Pap1-rPilt}. The interested reader may find
it in \cite[Sections 1.4 and 1.5]{MotDSc} (see also \cite{TMF1993}).
Here we only remark that the derivation is rather straightforward
being based directly on the representations \eqref{Pap1-16} or
\eqref{Pap1-16p} for the $T$-matrix.

The most important consequence of the representations
\eqref{Pap1-18}--\eqref{Pap1-rPilt} is the fact that all energy
singularities of the $T$-matrix, scattering matrices, and resolvent
on an unphysical sheet $\Pi_\ell$, differing of those in the
physical sheet, are just the singularities of the inverse truncated
scattering matrix $[s_\ell(z)]^{-1}$ (or, and this is the same, the
ones of its transpose $[s^\dagger_\ell(z)]^{-1}$). This means that
\begin{equation*}
\begin{array}{c}
\mbox{\textit{resonances on sheet $\Pi_\ell$ correspond exactly
to the points $z$}}\\
\mbox{\textit{on the physical sheet where the operator $s_\ell(z)$
has eigenvalue zero,}}
\end{array}
\eqno(\mathrm{R})
\end{equation*}
i.e. the resonances on $\Pi_\ell$ are those energies $z$ on $\Pi_0$
where equation
\begin{equation}
\label{Pap1-slA0} s_\ell(z)\mathcal{A}=0
\end{equation}
has a non-trivial solution $\mathcal{A}\neq0$ in the sum Hilbert
space $ \mathfrak{G} $ given by \eqref{fG}.

Eigenvectors of the truncated scattering matrices $s_\ell(z)$
associated with resonances have a quite transparent physical
meaning. Assume that $z$ is a resonance on the unphysical sheet
$\Pi_\ell$. This implies that for the same energy $z$ on the
physical sheet $\Pi_0$ equation \eqref{Pap1-slA0} has a solution
$\mathcal{A}\neq0$, $\mathcal{A}=(\mathcal{A}_1,\mathcal{A}_2,
\ldots,\mathcal{A}_m)^\dagger$. Clearly, the components
$\mathcal{A}_\alpha$ of the vector $\mathcal{A}$ are non-zero only
for the channels $\alpha$ such that $l_\alpha\neq0$. Taking into
account that \eqref{Pap1-slA0} can be written in the equivalent form
\begin{equation}
\label{Pap1-12p} \mathcal{A}=-\widetilde{L}Jt(z)J^\dagger
LA(z)\mathcal{A},
\end{equation}
this conclusion follows from equality
$$
(I-\widetilde{L})\mathcal{A}=0.
$$
Notice that the latter holds since
$\widetilde{L}(I-\widetilde{L})=0$.

Along with the vector $\mathcal{A}$ we also consider an ``extended''
vector $\widetilde{\mathcal{A}}$ that is obtained of $\mathcal{A}$
as a result of replacing the projection $\widetilde{L}$ on the
right-hand side of \eqref{Pap1-12p} with the identity operator, i.e.
\begin{equation}
\label{Pap1-12pp} \widetilde{\cal A}=-J t(z)J^\dagger LA(z) {\cal
A}.
\end{equation}
Clearly, $\mathcal{A}=\widetilde{L}\widetilde{\mathcal{A}}$.

We claim that up to scalar factors the components
$\widetilde{\mathcal{A}}_1(\,\widehat{{\bm{k}}}_1)$,
$\widetilde{\mathcal{A}}_2(\,\, \widehat{{\bm{k}}}_2)$, $\ldots$,
$\widetilde{\mathcal{A}}_m(\,\widehat{{\bm{k}}}_m)$ of the
eigenvector $\widetilde{\mathcal{A}}$ make sense of the breakup
amplitudes of the corresponding resonance state in channels 1,
2,$\ldots$, and $m$, respectively. In particular, these amplitudes
determine angular dependence of coefficients at the spherical waves
in the asymptotics of the channel components of the resonant
solution to the Schr\"odinger equation in coordinate representation.

To give some details, let us denote by $h_0^\#$ and $v^\#$ the
coordinate-space version (Fourier transform) of the operators $h_0$
and $v$, respectively. Namely, let
$$
h_0^\#=\mathop{\mathrm{diag}}(\lambda_1-\Delta_{{\bm{x}}_1},
\lambda_2-\Delta_{{\bm{x}}_2},\ldots,\lambda_m-\Delta_{{\bm{x}}_m}),
$$
where $\Delta_{\bm{x}_\alpha}$, $\alpha=1,2\ldots,m,$ stands for the
Laplacian in variable ${\bm{x}}_\alpha\in\mathbb{R}^{n_\alpha}$.

In the statement below we restrict ourselves to the case where
absolute values of the unphysical-sheet indices corresponding to the
even-dimensional channels are less than or equal unity, i.e. we
assume that if $n_\alpha$ is even then $|l_\alpha|\leq 1$. Recall
that if $n_\alpha$ is odd then automatically $l_\alpha=0$ or
$l_\alpha=1$.
\smallskip

\noindent{\bf Lemma 2}. {\it Assume that $z$ is a resonance on an
unphysical sheet $\Pi_\ell$ with multi-index
$\ell=(\ell_1,\ell_2,\ldots,\ell_m)$ such that $|\ell_\alpha|\leq 1$
for all $\alpha=1,2,\ldots,m$. Let ${\cal A}\in\mathfrak{G}$ be a
non-zero solution to equation \eqref{Pap1-slA0} for the same energy
$z$ but belonging to the physical sheet. Then for this $z$ the
Schr\"odinger equation
\begin{equation}
\label{Pap1-SchrCoor}
\left(h_0^\# +{v}^{\#}\right)\psi^{\#}=z\psi^{\#}
\end{equation}
has a non-zero {\rm(}resonant{\rm)} solution $\psi^{\#}_{\rm
res}=(\psi^{\#}_{\rm res,1},\psi^{\#}_{\rm res,2},
\ldots,\psi^{\#}_{{\rm res},n})^\dagger$ whose components
$\psi^{\#}_{\rm res,\alpha}(x_\alpha)$ for $\ell_\alpha\neq 0$
possess exponentially increasing asymptotics,
\begin{equation}
\label{Pap1-Gamow}
\psi^{\#}_{\rm res,\alpha}({\bm{x}}_\alpha)\mathop{=}
\limits_{{\bm{x}}_\alpha\rightarrow\infty}
C_\alpha(z,\ell_\alpha)\bigl({\cal
A}_\alpha(\,-\widehat{{\bm{x}}}_\alpha)+o(1)\bigr)
\dfrac{{\rm
e}^{-\mathrm{i}\sqrt{z-\lambda_\alpha}|{\bm{x}}_\alpha|}}
{|{\bm{x}}_\alpha|^{(n_\alpha-1)/2}},
\end{equation}
while for $\ell_\alpha=0$ their asymptotics is exponentially
decreasing,
\begin{equation}
\label{Pap1-Gamow0}
\psi^{\#}_{\rm res,\alpha}({\bm{x}}_\alpha)\mathop{=}
\limits_{{\bm{x}}_\alpha\rightarrow\infty}
C_\alpha(z,\ell_\alpha)\bigl(\widetilde{\mathcal{A}}_\alpha
(\,\widehat{{\bm{x}}}_\alpha) +o(1)\bigr)
\dfrac{{\rm
e}^{+\mathrm{i}\sqrt{z-\lambda_\alpha}|{\bm{x}}_\alpha|}}
{|{\bm{x}}_\alpha|^{(n_\alpha-1)/2}},
\end{equation}
where $\widetilde{\mathcal{A}}(\,\widehat{{\bm{k}}}_\alpha)$ stand
for the corresponding components of the extended vector
\eqref{Pap1-12pp} and
\begin{equation}
\label{Pap-Cas}
C_\alpha(z,\ell_\alpha)=\sqrt{\dfrac{\pi}{2}}\,\,\mathrm{e}^{\mathrm{i}
\frac{(n_\alpha-3)(2\ell_\alpha-1)}{4}\pi}
(z-\lambda_\alpha)^{\frac{n_\alpha-3}{4}}
\end{equation}
For the function $(z-\lambda_\alpha)^{\frac{n_\alpha-3}{4}}$ on the
right-hand side of \eqref{Pap-Cas} the main branch is chosen.}

Complete proof of this statement may be found in
\cite[Section\,1.6]{MotDSc}.

The functions $\psi^{\#}_{\rm res, \alpha}(x_\alpha)$ taken
altogether form the Gamov vector corresponding to the
resonance energy $z$ (see, e.g. \cite{BohmQM,Newton}). Just
asymptotic formulas (\ref{Pap1-Gamow}) and (\ref{Pap1-Gamow0}) prove
that the functions ${\cal A}_\alpha(\,\widehat{k}_\alpha)$,
$\ell_\alpha\neq 0$, and
$\widetilde{\mathcal{A}}_\alpha(\,\widehat{k}_\alpha)$,
$\ell_\alpha=0$, make sense of the breakup amplitudes describing decay
of the resonant state along open and closed channels, respectively.

\section{Three-body problem}
\label{Sec-3-body}

In this section we give a sketch of our results on the structure of
the $T$-matrix, scattering matrices, and Green function on
unphysical energy sheets in the three-body problem. For detail
exposition of this material see Refs. \cite{MN1997} or
\cite{MotDSc}.

Let $H_0$ be the three-body kinetic energy operator in the
center-of-mass system. Assume that there are no three-body forces
and thus the total interaction reads $V=v_1+v_2+v_3$ where
$v_\alpha$, $\alpha=1,2,3$, are the corresponding two-body
potentials having just the same properties as in Sec.
\ref{Sec-2-body}.

The best way to proceed in the three-body case is to work with the
Faddeev components~\cite{Faddeev1963}
$$
M_{\alpha\beta}=\delta_{\alpha\beta}v_{\alpha}-v_{\alpha}
G(z)v_{\beta} \quad (\alpha,\beta=1,2,3)
$$
of the $T$-operator $T(z)=V-VG(z)V$ where $G(z)$ denotes the
resolvent of the total Hamiltonian $H=H_0+V$. The components
$M_{\alpha\beta}$ satisfy the Faddeev equations
\begin{equation}
\label{F}
M_{\alpha\beta}(z)=\delta_{\alpha\beta}\mathbf{t}_\alpha(z)-
\mathbf{t}_\alpha(z) G_0(z)\sum_{\gamma\neq\alpha}M_{\gamma\beta}(z)
\end{equation}
with $G_0(z)=(H_0-z)^{-1}$ and
$$
\mathbf{t}_\alpha(P,P',z)=t_\alpha({\bm{k}}_\alpha,
{\bm{k}}'_\alpha,z-{\bm{p}}_\alpha^2)
\delta({\bm{p}}_\alpha-{\bm{p}}'_\alpha)
$$
where ${\bm{k}}_\alpha,{\bm{p}}_\alpha$ denote the corresponding
reduced Jacobi momenta (see \cite{MN1997} for the precise definition
we use) and $P=({\bm{k}}_\alpha,{\bm{p}}_\alpha)\in\mathbb{R}^6$ is
the total momentum.

Assume that any of the three two-body subsystems has only one bound
state with the corresponding energy $\varepsilon_\alpha<0$,
$\alpha=1,2,3$. Assume in addition that all of these three binding
energies are different. It is easy to see that the thresholds
$\varepsilon_1$, $\varepsilon_2$, $\varepsilon_3$, and 0 are
associated with particular Cauchy type integrals in the integral
equations \eqref{F}. By Lemma 1 the two-body thresholds
$\varepsilon_\alpha$ appear to be square-root branching points while
the three-body threshold 0 is the logarithmic one. In order to
enumerate the unphysical sheets we introduce the multi-index
$\ell=(\ell_0,\ell_1,\ell_2,\ell_3)$ with
$\ell_0=\ldots,-1,0,1,\ldots$ and $\ell_\alpha=0,1$ if
$\alpha=1,2,3$. Clearly, only encircling the two-body thresholds one
arrives at seven unphysical sheets. The three-body threshold
generates infinitely many unphysical sheets. (There might also be
additional branching points on the unphysical sheets, in particular
due to two-body resonances.)

It turns out that the analytically continued Faddeev equations
\eqref{F} can be explicitly solved in terms of the matrix
$M=\{M_{\alpha\beta}\}$ itself taken only on the physical sheet,
just like in the case of the two-body $T$-matrix in Sec.
\ref{Sec-2-body} and multichannel $T$-matrix in Sec. \ref{Sec-mult}.
The result strongly depends, of course, on the unphysical sheet
$\Pi_{\ell}$ concerned. More precisely, the resulting representation
reads as follows
\begin{equation}
\label{MPi} M|_{\Pi_{\ell}}=M+Q_M
\,L\,S_{\ell}^{-1}\,\widetilde{L}\,\widetilde{Q}_M.
\end{equation}
In the particular case we deal with, $L$ and $\widetilde{L}$ are
$4\times4$ diagonal scalar matrices of the form
$L=\mathop{\mathrm{diag}}({\ell}_0, {\ell}_1,$ ${\ell}_2,{\ell}_3)$
and $\widetilde{L}=\mathop{\mathrm{diag}}
(|{\ell}_0|,{\ell}_1,{\ell}_2,{\ell}_3)$, respectively;
$S_{\ell}(z)=I+\widetilde{L}(S(z)-I)L$ is a truncation of the total
scattering matrix $S(z)$ and the entries $Q_M$, $\widetilde{Q}_M$
are explicitly written in terms of the half-on-shell kernels of $M$
(see formula (7.34) of \cite{MN1997}). {}From \eqref{MPi} one also
derives explicit representations for $G(z|_{\Pi_{\ell}})$ and
$S(z|_{\Pi_{\ell}})$ similar to those of \eqref{Pap1-rPil} and
\eqref{Pap1-18}, respectively.

Thus, to find resonances on the sheet $\Pi_{\ell}$ one should simply
look for the zeros of the truncated scattering matrix $S_{\ell}(z)$,
that is, for the points $z$ on the physical sheet where equation
$S_{\ell}(z)\mathcal{A}=0$ has a non-trivial solution $\mathcal{A}$.
The vector $\mathcal{A}$ will consist of amplitudes of the resonance
state to  breakup into the various possible channels. Within such an
approach one can also find the three-body virtual states.

In order to find the amplitudes involved in $S_{\ell}$, one may
employ any suitable method, for example the one of Refs.
\cite{CPC2000,YaF1999,YaF1997} based on the Faddeev differential
equations. In these works the approach we discuss has been
successfully applied to several three-body systems. In particular,
the mechanism of emerging the Efimov states in the $^4$He trimer has
been studied \cite{CPC2000,YaF1999}.

\acknowledgements The author kindly acknowledges support of this
work by the Russian Foundation for Basic Research and the
Deu\-ts\-che Forschungsgemeinschaft (DFG).

\end{document}